\documentclass[twocolumn,secnumarabic,amsmath,amssymb, nobibnotes, aps, prl,superscriptaddress]{revtex4-1}

\pdfoutput=1
\RequirePackage{lineno}
\usepackage{hyperref}
\usepackage{graphicx}
\usepackage{bm}
\usepackage{dcolumn}
\usepackage{braket}
\usepackage{epstopdf}
\usepackage{color}

\begin{document}

\title{Optically and electrically controllable adatom spin-orbital dynamics in transition metal dichalcogenides}%

\author{B. Shao}%
\email{bin.shao@uni-bremen.de}
\affiliation{Bremen Center for Computational Materials Science, Universit\"at Bremen, Am Fallturm 1, 28359 Bremen, Germany}
\affiliation{Institut f\"ur Theoretische Physik, Universit\"at Bremen, Otto-Hahn-Allee 1, 28359 Bremen, Germany}

\author{M. Sch\"uler}
\affiliation{Bremen Center for Computational Materials Science, Universit\"at Bremen, Am Fallturm 1, 28359 Bremen, Germany}
\affiliation{Institut f\"ur Theoretische Physik, Universit\"at Bremen, Otto-Hahn-Allee 1, 28359 Bremen, Germany}

\author{G. Sch\"onhoff}
\affiliation{Bremen Center for Computational Materials Science, Universit\"at Bremen, Am Fallturm 1, 28359 Bremen, Germany}
\affiliation{Institut f\"ur Theoretische Physik, Universit\"at Bremen, Otto-Hahn-Allee 1, 28359 Bremen, Germany}

\author{T. Frauenheim}
\affiliation{Bremen Center for Computational Materials Science, Universit\"at Bremen, Am Fallturm 1, 28359 Bremen, Germany}

\author{G. Czycholl}
\affiliation{Institut f\"ur Theoretische Physik, Universit\"at Bremen, Otto-Hahn-Allee 1, 28359 Bremen, Germany}

\author{T. O. Wehling}
\affiliation{Bremen Center for Computational Materials Science, Universit\"at Bremen, Am Fallturm 1, 28359 Bremen, Germany}
\affiliation{Institut f\"ur Theoretische Physik, Universit\"at Bremen, Otto-Hahn-Allee 1, 28359 Bremen, Germany}

\date{\today}%
\begin{abstract}
We analyze the interplay of spin-valley coupling, orbital physics and magnetic anisotropy taking place at single magnetic atoms adsorbed on semiconducting transition metal dichalcogenides, MX$_2$ (M = Mo, W; X = S, Se). Orbital selection rules turn out to govern the kinetic exchange coupling between the adatom and charge carriers in the MX$_2$ and lead to highly orbitally dependent spin-flip scattering rates, as we illustrate for the example of transition metal adatoms with $d^9$ configuration. Our \textit{ab initio} calculations suggest that $d^9$ configurations are realizable by single Co, Rh, or Ir adatoms on MoS$_2$, which additionally exhibit a sizable magnetic anisotropy. We find that the interaction of the adatom with carriers in the MX$_2$ allows to tune its behavior from a quantum regime with full Kondo screening to a regime of "Ising spintronics" where its spin-orbital moment acts as classical bit, which can be erased and written electronically and optically. 
\end{abstract}

\maketitle

%\section{Introduction}

Transition metal adatoms on surfaces provide ideal model systems for fundamental studies of quantum many-body phenomena ranging from magnetism\cite{gambardella_giant_2003,rau_reaching_2014,baumann_origin_2015} and Kondo physics\cite{manoharan_quantum_2000,otte_role_2008,ternes_spectroscopic_2009,khajetoorians_tuning_2015} to topological states of matter\cite{liu_magnetic_2009,wray_topological_2011,honolka_-plane_2012} and Majorana modes\cite{vazifeh_self-organized_2013,nadj-perge_observation_2014,peng_strong_2015}. Moreover, these systems are promising as ultimately miniaturized building blocks of spintronic devices and logic gates. Particularly recent advances in scanning tunneling microscopy lead to enormous progress in the probing and manipulation of these systems including writing, reading, and processing of information from atomic scale bits via e.g. spin-transfer torques\cite{loth_controlling_2010,khajetoorians_current-driven_2013} and spin-polarized spectroscopy techniques\cite{heinrich_single-atom_2004,meier_revealing_2008}. 

In all of these cases, the coupling between adatom and substrate is central to determine the magnetic properties of the system. Thus, changes in the quantum state of the substrate can directly affect the adatom magnetism, as studies of superconducting substrates demonstrated~\cite{yazdani_probing_1997,heinrich_protection_2013}. In the light of time-dependent phenomena, substrates which allow for ultrafast manipulation of their electronic states by electronic or optical means are particularly interesting but actual realizations have been lacking so far.

In this letter, we show that strong spin-valley coupling and peculiar orbital physics make monolayers of transition metal dichalcogenides (TMDCs), MX$_2$ (M = Mo, W; X = S, Se) ideal substrates in this context. MX$_2$ materials allow for ultrafast optical control of their electronic states and for charge doping by external gates, which turn out to provide control over spin-flip scattering of transition metal (TM) adatoms on a monolayer MX$_2$. We illustrate this result based on \textit{ab initio} simulations of single Co, Rh, or Ir adatoms on MoS$_2$ and a generic model Hamiltonian description. Our calculations show, that these magnetic adatoms exhibit a doublet ground state which is separated from excited states by a sizable magnetic anisotropy $>10$\,meV and realizes an "Ising" spin-orbital moment. We analyze the kinetic exchange scattering of adatom and substrate electrons, and demonstrate that the adatom behavior can be tuned from a regime of "Ising spintronics" where its spin-orbital moment acts as classical bit, which can be manipulated electronically and optically, to a quantum regime of full Kondo screening.

\begin{figure}%figure1
	\includegraphics[width=\linewidth]{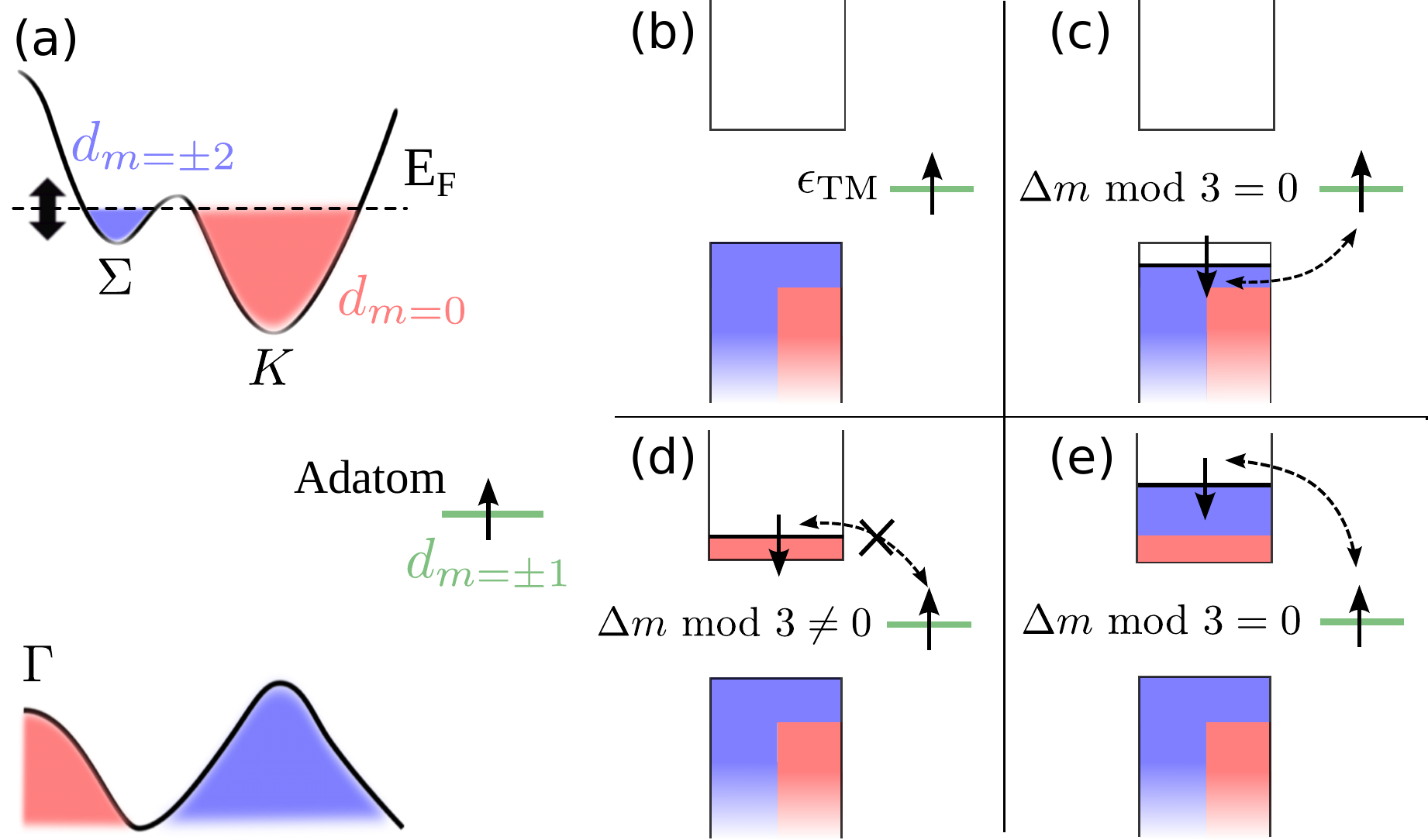}
	\caption{\label{fig1}(Color online) Illustration of the spin-orbital selective coupling between the TM adatom and charge carriers in MX$_2$. (a) A typical MX$_2$ band structure and the characters of M atom $d$ orbitals are denoted by different colors. Due to the orbital selection rule, $\Delta m\,\text{mod}~3=0$, the spin-flip scattering sensitively depends on the level of charge doping. In the cases of (b) neutral and (d) moderate electron doping, the spin state of adatom is preserved due to a lack of effective elastic scattering channels. However, if electrons at the Fermi level in MX$_2$ carry $d_{m=2}$ orbital character as in the cases of (c) moderate hole doping and (e) relatively high electron doping, the spin-flip scattering becomes possible and the spin state is eventually fully screened. Due to spin-orbit coupling the spin and the orbital moment of the adatom are flipped simultaneously in the scatterings.}
\end{figure}

The choice of MX$_2$ as substrate concerns two aspects: First, there are adsorption sites with uniaxial symmetry ($C_{3v}$) on the surface of MX$_2$, i.e. the top of M atoms (M-top), the top of X atoms (X-top), and the site above the middle of M-X hexagons (hollow). Uniaxial symmetry is crucial for TM adatom to retain a large orbital moment and consequently yields a sizable magnetic anisotropy~\cite{rau_reaching_2014,baumann_origin_2015}. 
Moreover, the symmetry determines the hybridization between TM and M atoms. Hence, only those $d$ orbitals of TM and M atoms with matching symmetry under the operation of the adsorption site point group can couple to each other, providing an additional degree of freedom to control the spin-flip scattering. Second, we can easily select the spin and orbital character of charge carriers in MX$_2$. As shown in Fig.~\ref{fig1}(a), the band-edges of MX$_2$ result from different valleys which are predominately stemming from different $d$ orbitals of the M atoms\cite{cappelluti_tight-binding_2013, liu_three-band_2013, rosner_phase_2014}.

For example, the lowest conduction-band (CB) in the $K$ valley carries mostly $d_{m=0}$ character, while we have mainly $d_{m=\pm2}$ in the $\Sigma$ valley; in the upmost valence-band (VB) we have mostly $d_{m=0}$ in the $\Gamma$ valley and $d_{m=\pm2}$ in the $K$ valley. Here $m$ is the quantum number of the orbital momentum's $z$ component. Given the energy separation between the minima/maxima in CBs/VBs, charge doping thus selects the orbital character of carriers at the Fermi level in MX$_2$. Due to the $C_{3v}$ symmetry of adsorption sites, the coupling $d_m$ orbitals from M atoms and TM adatoms need to follow an orbital selection rule, $\Delta m~\text{mod}~3=0$. Therefore, spin-flip scattering will strongly depend on the doping level. Specifically, if the valence hole of a $d^9$ adatom resides in a $d_{m=\pm1}$ state [Fig.~\ref{fig1}(a)], spin-flip scattering by charge carriers in MX$_2$ is suppressed for the undoped system due to the absence of any carriers [Fig.~\ref{fig1}(b)] but also for the moderately electron doped case due to the absence of symmetry matched carriers [Fig.~\ref{fig1}(d)]. However, once carriers with $d_{m=\pm2}$ orbital character are available for scattering at the Fermi level, e.g., in the cases of moderate hole doping [Fig.~\ref{fig1}(c)] and relatively high electron doping [Fig.~\ref{fig1}(e)], an effective channel for spin-flip opens and the adatom spin might be even fully screened by the charge carriers. More significantly, because of the so-called spin-valley coupling, TMDCs with broken inversion symmetry, e.g. MX$_2$ monolayers, allow for optical selecting the spin state of the excited carriers\cite{xiao_coupled_2012, suzuki_valley-dependent_2014,xu_spin_2014}. This optically determined spin state can be further coupled to the adatom spin via spin-flip scattering, providing a mechanism of optical orientation of the adatom spin. The orbital and spin selection rules of spin-flip scattering present the basis for the ultrafast control over the magnetic adatom degrees of freedom suggested in this letter.

%\section{Model}

To investigate this scenario quantitatively, we firstly study the ground state of the TM adatom subject to the crystal field with $C_{3v}$ symmetry and spin-orbit coupling (SOC), which can be described by the effective Hamiltonian, 
\begin{equation}\label{eq1}
H = H_{\text{CF}}+\sum_{i}\lambda {{\bm{l}}_i} \cdot {\bm{s}_i}.
\end{equation}
Here $H_{\text{CF}}$ is the crystal field Hamiltonian, $\lambda$ is the spin-orbit constant, $\bm{l}_i$ and $\bm{s}_i$ are the orbital and spin angular momentum vector of an electron, respectively, the sum over $i$ runs over all the electrons. Because we are interested in a $d^9$ configuration the atomic state is a single Slater determinant. In the basis of the five $d$ orbitals ($d_{m=0,\pm1,\pm2}$), $H_{\text{CF}}$ is given by a $5\times5$ matrix. Due to the three fold rotation axis of the crystal field, the matrix element $\bra{d_{m_i}}H_{\text{CF}} \ket{d_{m_j}}$ is non-vanishing only when $\vert m_i - m_j \vert \mod 3 = 0$. Those elements include $\bra{d_{m}}H_{\text{CF}} \ket{d_{m}}$, which is the energy of $d_{m}$ orbital ($\epsilon_m$), and, $\bra{d_{-1}}H_{\text{CF}} \ket{d_{+2}}$ and $\bra{d_{+1}}H_{\text{CF}} \ket{d_{-2}}$ labeled by $c_{-3}$ and $c_{+3}$, respectively. In absence of magnetic fields, time reversal symmetry further implies $\epsilon_{-2}=\epsilon_{+2}$ and $\epsilon_{-1}=\epsilon_{+1}$. Then, $H_{\text{CF}}$ reads
\begin{equation}\label{eq2}
H_{\text{CF}}=
\begin{pmatrix}
 \epsilon_{-2}  &  0 & 0   & c^{\star}_{+3}   & 0   \\
 0   & \epsilon_{-1}  & 0   & 0   & c_{-3}   \\
 0   & 0   & \epsilon_{ 0}  & 0   & 0   \\
 c_{+3}   & 0  & 0   & \epsilon_{+1}  & 0 \\
 0   & c^{\star}_{-3}   & 0   & 0 & \epsilon_{+2}
\end{pmatrix}.
\end{equation}
The crystal field with $C_{3v}$ symmetry splits the five $d$ orbitals into two doublets, $E1$ (mostly $d_{\pm1}$) and $E2$ (mostly $d_{\pm2}$), and a singlet $A1$ ($d_0$). By diagonalizing (\ref{eq1}), we can obtain the spin-orbit state of the $d^9$ configuration.

In order to show that the $d^9$ configuration is a realistic scenario for several single TM adatoms on MX$_2$ and to obtain reasonable parameters in (\ref{eq2}) we performed \textit{ab initio} calculations in the framework of density functional theory (DFT)~\cite{kohn_self-consistent_1965}. We employed a $5 \times 5$ MoS$_2$ supercell with one Co, Rh, or Ir adatom using Perdew-Burke-Ernzerhof parameterization~\cite{perdew_generalized_1996} of generalized gradient approximation (GGA) functional and the projector augmented wave~\cite{blochl_projector_1994, kresse_ultrasoft_1999} as implemented in the VASP package~\cite{kresse_efficient_1996}. All the atomic structures were fully relaxed before calculating their electronic structures. 

\begin{figure}%figure2
	\includegraphics[width=\linewidth]{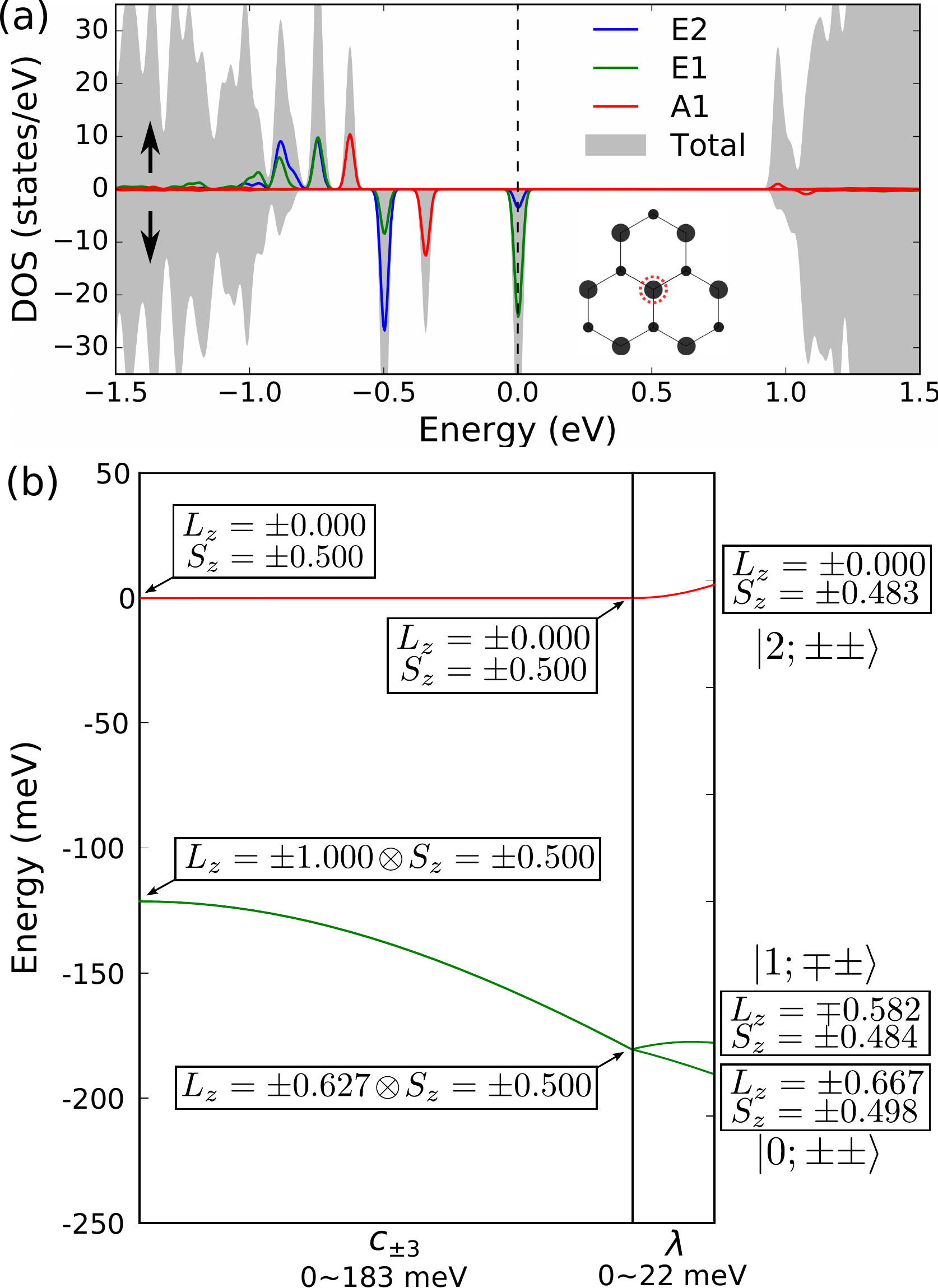}
	\caption{\label{fig2}(Color online) (a) Projected spin-polarized density of states of a Co adatom on a Mo-top site of a MoS$_2$ monolayer obtained with the GGA method. The up (down) arrow denotes the majority (minority) spin. The Fermi level is at 0 eV. The red, green, and blue lines refer to the $A_1$, $E_1$, and $E_2$ states, respectively. The dashed circle in the insert indicates the Mo-top site of the MoS$_2$, where the Co adatom is located, and the large (small) dots represent Mo (S) atoms. (b) Energy level diagram obtained by diagonalizing the Hamiltonian (\ref{eq1}) as a function of the crystal field parameter $c_{\pm3}$ and the spin-orbital coupling $\lambda$. The $L_z$ and $S_z$ values are the expectation value of orbital and spin angular momentum along $z$ axis, respectively. The states are denoted by labels $\ket{n=\{0,1,2,...\};L_zS_z=\{\pm\pm,\mp\pm\}}$. Here, $n$ is assigned to numbering the doublets in order of increasing energy and "$\pm\pm$" ("$\mp\pm$") denotes the orbital and spin momentum directions are parallel (antiparallel) to each other.}
\end{figure}

We have considered three adsorption sites, Mo-top, S-top, and hollow. Our results indicate that the Mo-top site is energetically favored by all three adatoms, and the other two sites become energetically more unfavorable from Co to Ir (see Table S1 in the supplemental material). The spin-polarized GGA calculations predict an electronic configuration close to $d^9$ for all the adatoms, in which one hole resides in $E1$ state which has predominantly $d_{\pm1}$ orbital character. Here, without loss of generality we take Co on MoS$_2$ as a representative example to be discussed in the following. As shown in the projected density of states (PDOS) [(Fig.~\ref{fig2}(a)], the $d$ orbital order of minority spin in energy is $E2 < A1 < E1$ with hybridization between $E2$ and $E1$, which is in agreement with the allowed off-diagonal terms of the crystal field in (\ref{eq2}). To investigate the impacts of on-site Coulomb repulsion on the results, we also carried out GGA+$U$ calculations for the Co adatom~\cite{liechtenstein_density-functional_1995}. The GGA+$U$ calculations do not change the orbital ordering and occupation. Only for large $U\geqslant7.0$ eV ($J=0.9$ eV) $E1$ is shifted into the CBs.

We extract the following crystal field parameters of $H_{\text{CF}}$ from the DFT calculations of Co on MoS$_2$: $\epsilon_{\vert m \vert =2} = -0.287$ eV, $\epsilon_{\vert m \vert =1} = 0.130$ eV, $\epsilon_{m=0}=0$ eV, and $\vert c_{\pm3} \vert = 0.183$ eV. Setting the spin-orbit constant to $\lambda=22$ meV~\cite{rau_reaching_2014}, the spin-orbit eigenstates of Co are obtained by diagonalizing (\ref{eq1}) and are labelled by $\ket{n=\{0,1,2,...\};L_zS_z=\{\pm\pm,\mp\pm\}}$. Here, $n$ is assigned to numbering the states in order of increasing energy. The  $L_z$ and $S_z$ refer to the orbital and spin angular momentum along $z$ axis, respectively, and "$\pm\pm$" ("$\mp\pm$") denotes the orbital and spin momentum directions are parallel (antiparallel) to each other. In Fig.~\ref{fig2}(b) we illustrate the evolution of the lowest eigenstates under the variation of the crystal field parameter $c_{\pm3}$ and the spin-orbit coupling $\lambda$. It shows that $c_{\pm3}$ perturbs the lowest quadruplet, partially quenching its $L_z$. The SOC splits the quadruplet into two doublets with different relative alignments of the spin and orbital moments. The energy separation between the ground state $\ket{0;\pm\pm}$ and the first excited state $\ket{1;\mp\pm}$ is about 12.0 meV. I.e. we have an effective out-of-plane magnetic anisotropy on the order of 12 meV for the coupled spin-orbital moment of the Co adatom, which is essentially comparable to the case of Co on Pt (111)~\cite{gambardella_giant_2003, meier_revealing_2008}.

In order to investigate the stability of the spin-orbital moment, we inspect possible final states as resulting from pure spin-flip scattering if the initial state is one of the spin-orbital ground states $\ket{0;++}$ of the adatom. We estimate the probabilities of reaching the first excited state and the other degenerate ground state by calculating $\vert \bra{1;+-}S^{-}\ket{0;++} \vert^2$ and $\vert\bra{0;--}S^{-}\ket{0;++}\vert^2$, where $S^{-}$ is the ladder operator to transit $S_z$ from "$+$" to "$-$". Our result shows that the former is about 0.98 and the latter is numerically 0, indicating the transition inside the ground state manifold is suppressed. For the former transition, we need to overcome the excitation energy from the ground state to the first excited state arising from SOC which is proportional to $\lambda L$~\cite{rau_reaching_2014}, where $L$ is the orbital moment magnitude. Given that the corresponding excitation energy is $\sim12$ meV for Co and that of 4d and 5d TM adatoms will be larger because of the stronger SOC, we can neglect the corresponding transition at sufficiently low temperatures. Thus, the spin-orbital moment of adatoms is essentially not perturbed by pure spin-flip scattering. However, carriers in MoS$_2$ can scatter with Co, disturbing its quantum state via hybridization. For example, the transition  $\bra{d^{\text{Co}}_{\pm1}}H_{\text{hyb}}\ket{d^{\text{Mo}}_{\mp2}}\bra{d^{\text{Mo}}_{\pm2}}H_{\text{hyb}}\ket{d^{\text{Co}}_{\mp1}}\neq 0$ which meets the orbital selection rule $\vert m_i - m_j \vert \mod 3 = 0$ can flip the spin and the orbital moment simultaneously [see Fig.\ref{fig1}(c), (e)] and provides a channel for elastic scattering.

%Eventually, we realize an "Ising" spin-orbital moment in this doublet ground state.

%Since MoS$_2$ is a semiconductor, the interaction between its electrons and Co is weak for the charge neutral case. 

%\subsection*{Spin lifetime}
To address the impact of interaction with charge carriers in MoS$_2$ on the spin-orbital moment of Co, we describe the system in terms of an Anderson impurity model (AIM)
\begin{equation}\label{eq3}
H = H_{\text{MoS}_2} + H_{\text{Co}} + H_{\text{hyb}},
\end{equation}
The first term corresponds to the charge carriers residing in MoS$_2$ and we employ a tight-binding (TB) Hamiltonian from Ref.~\cite{liu_three-band_2013} in the basis of three Mo $d$ orbitals ($m=0,\pm2$)\footnote{For convenience, the three $d$ orbitals are denoted by the $z$-component of the orbital momentum quantum number $m$. In the calculations, we employed their real forms $d_{z^2}$, $d_{xy}$ and $d_{x^2-y^2}$ as the basis, which can be obtained by the relation: $d_{z^2}=d_{m=0}$, $d_{xy}=-i2^{-1/2}(d_{m=2}-d_{m=-2})$, $d_{x^2-y^2}=2^{-1/2}(d_{m=2}+d_{m=-2})$.} relevant for the band electrons of MoS$_2$. For Co, we consider its five $d$ orbitals split by the crystal field and SOC as in (\ref{eq2}) and the local Coulomb interaction $U$. The third term describes the hybridization between the Co and MoS$_2$ monolayer and reads
%\begin{align}\label{eq4}
%H_{\text{hyb}}
%=\sum_{m_1m_2\sigma} V_{m_1m_2}d^{\dagger}_{m_1\sigma}c_{m_2\sigma\mathbf{R_{0}}}+ V^{\ast}_{m_2m_1}c^{\dagger}_{m_2\sigma\mathbf{R_{0}}}d_{m_1\sigma},
%\end{align}
\begin{equation}\label{eq4}
H_{\text{hyb}}
=\sum_{m_1,m_2,\sigma} V_{m_1m_2}d_{m_1,\sigma}c^{\dagger}_{m_2,\sigma}+ \text{H.c.},
\end{equation}
where $m_1$ and $m_2$ are the $z$ component of $d$ orbital quantum number of Co and Mo, respectively. $d_{m_1,\sigma}$ is the Fermi operator of Co $d$ electrons. $c_{m_2,\sigma}=\sum_{\mathbf{k}}c_{\mathbf{k},m_2,\sigma}$ is the Fermi operator of charge carriers in MoS$_2$ with orbital quantum number $m_2$ at the adsorption site $\mathbf{R_0 =0}$. $V_{m_1m_2}$ is the hybridization matrix element which can be non-zero if $\vert m_1 -m_2\vert~\text{mod}~3 = 0$. We obtain the hybridization matrix elements by fitting the hybridization function obtained from the model Hamiltonian to that obtained from our DFT calculations. (For details, see the supplemental material).
  
To describe the spin-orbital-flip scattering, which is included in the AIM (\ref{eq4}) only by virtual processes, directly, we reduce it to a kinetic exchange Hamiltonian via a Schrieffer-Wolff transformation~\cite{schrieffer_relation_1966},
\begin{equation}\label{eq5}
\begin{split}
V_{\text{ex}} =& \sum_{\mathbf{k},\mathbf{k'}}J_{\mathbf{k}\mathbf{k'}}\lbrace \tilde{S}^{+}c^{\dagger}_{\mathbf{k'},-2,\downarrow} c_{\mathbf{k},+2,\uparrow} + \tilde{S}^{-}c^{\dagger}_{\mathbf{k'},+2,\uparrow} c_{\mathbf{k},-2,\downarrow} \\
&+\tilde{S}_{z}(c^{\dagger}_{\mathbf{k'},+2,\uparrow}c_{\mathbf{k},+2,\uparrow}-c^{\dagger}_{\mathbf{k'},-2,\downarrow}c_{\mathbf{k},-2,\downarrow})\rbrace.\\
\end{split}
\end{equation}
Here, since the kinetic exchange Hamiltonian refers to the ground state manifold only, the operators $\tilde{S}_z$ and $\tilde{S}^{\pm}$ ($=\tilde{S}_{x}\pm i\tilde{S}_{y}$) are pseudo-spin operators. $\tilde{S}^{\pm}$ switches the state of Co between $\ket{0;--}$ and $\ket{0;++}$ and thus refers to simultaneous flipping of spin and orbital moments. $J_{\mathbf{k}\mathbf{k'}}$ represents the coupling constant of the kinetic exchange interaction, which relates to the parameters of AIM by
\begin{equation}\label{eq6}
J_{\mathbf{k}\mathbf{k'}} = V^{\ast}_{m_1m_2}V_{m_2m_1}\lbrace \frac{1}{U+\epsilon_{\text{d}}-\epsilon_{\mathbf{k'}}}+\frac{1}{\epsilon_{\mathbf{k}}-\epsilon_{\text{d}}}\rbrace,
\end{equation} 
where $\epsilon_{\text{d}}$ and $\epsilon_{\mathbf{k}}$ are the eigenvalues of the matrix representations of $H_{\text{Co}}$ and $H_{\text{MoS}_2}$, respectively. The spin lifetime $\tau(\epsilon_{\mathbf{k}})$ of Co is obtained by calculating the scattering rate of a carrier of $\text{MoS}_2$ transiting from a state $\ket{\mathbf{k},+2,\uparrow}$ to $\ket{\mathbf{k}',-2,\downarrow}$ due to the kinetic exchange interaction. Here, $\ket{\mathbf{k},+2,\uparrow}$ refers to a spin up carrier with $d_{m=+2}$ orbital character. The rate is given by 
\begin{align}\label{eq7}
\begin{split}
W(\mathbf{k}\uparrow\rightarrow\mathbf{k}'\downarrow) =& \frac{2\pi}{\hbar N^2_{\text{k}}} \vert \bra {\mathbf{k'},-2,\downarrow}V_{\text{ex}} \ket{\mathbf{k},+2,\uparrow}\vert ^2  \\
&\cdot\delta(\epsilon_{\mathbf{k}}-\epsilon_{\mathbf{k'}})f(\epsilon_{\mathbf{k}})(1-f(\epsilon_{\mathbf{k'}})),
\end{split}
\end{align}
where $1/N^2_{\text{k}}$ is a normalization factor and $N_{\text{k}}$ is the total number of $k$ points in Brillouin zone. $f(\epsilon)$ is the occupation number of the initial and final states of the electrons in MoS$_2$ and given by the Fermi distribution function in the thermal equilibrium. At the end, the spin lifetime $\tau(\mu)$ is derived as the inverse of the sum of (\ref{eq7}) over $\mathbf{k}$ and $\mathbf{k'}$ with $\epsilon_{\mathbf{k}'}=\epsilon_{\mathbf{k}}=\mu$, where $\mu$ is the chemical potential in MoS$_2$.
%\begin{equation}\label{eq8}
%\frac{1}{\tau(\mu)}=\frac{2\pi}{\hbar }\rho_{m=\pm2}(\mu)\vert V_{12}\vert^2 \lbrace \frac{1}{U+\epsilon_{\text{d}}-\mu}+\frac{1}{\mu-\epsilon_{\text{d}}}\rbrace.
%\end{equation}
%\begin{equation}\label{eq8}
%\frac{1}{\tau(\mu)}=\frac{2\pi}{\hbar }\rho(\mu)\vert\bra{\mathbf{k}',-2,\downarrow}V_{\text{ex}}\ket{\mathbf{k},+2,\uparrow}\vert^2_{\epsilon_{\mathbf{k}'}=\epsilon_{\mathbf{k}}=\mu}.
%\end{equation}
%Here, $\rho_{m=\pm2}(\mu)$ is the density of states projected to Mo $d_{m=\pm2}$ orbitals, and we assume that the hybridization matrix element between Co $d_{m=\pm1}$ orbitals and Mo $d_{m=\mp2}$ orbitals, $V_{12}$, is $\mathbf{k}$ independent.

\begin{figure}%figure3
 \includegraphics[width=\linewidth]{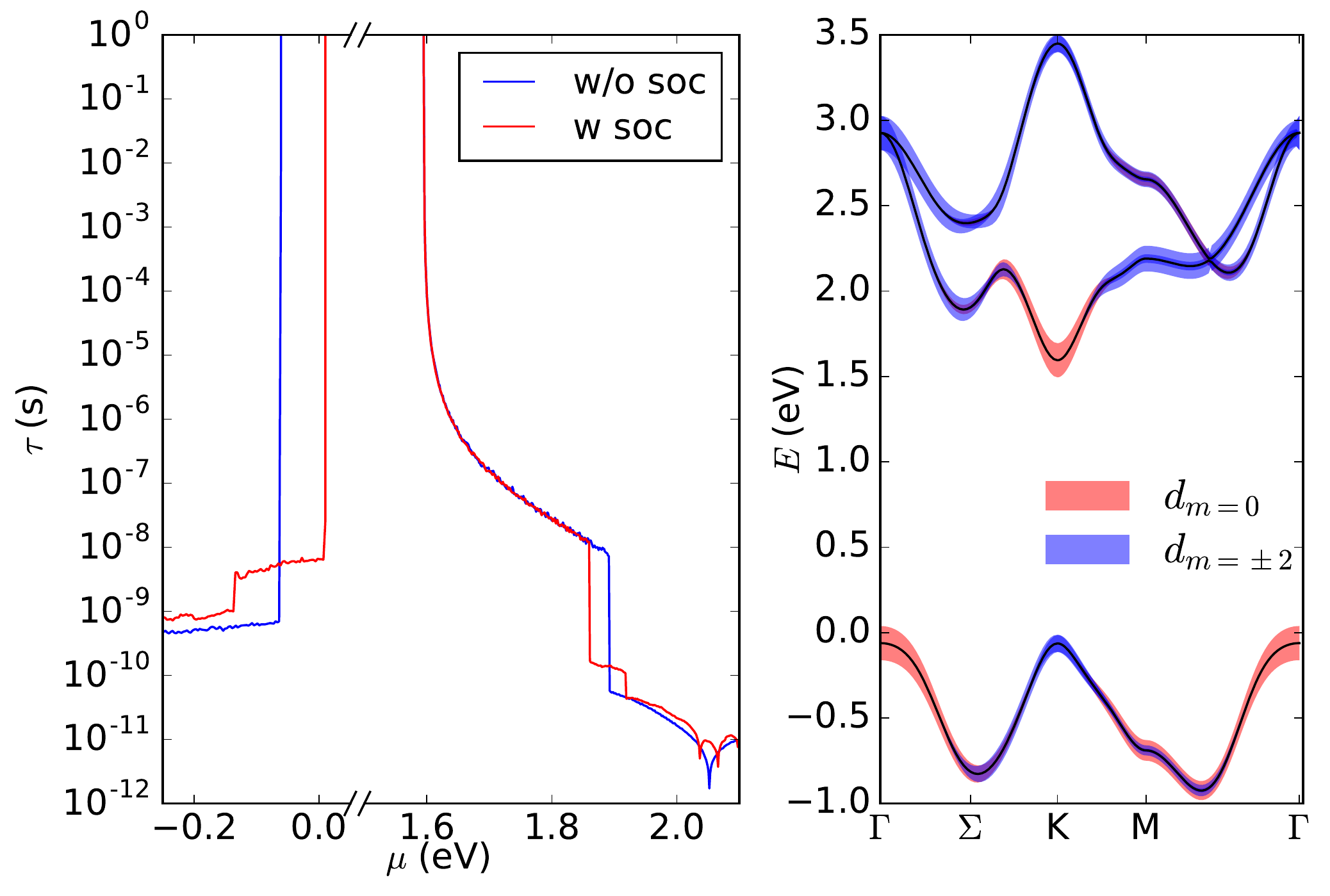}
 \caption{\label{fig3}(Color online) The spin lifetime $\tau$ as a function of chemical potential $\mu$ (left) and the TB band structure of MoS$_2$ (right). In the left panel, the blue and red line refer to the result without and with SOC, respectively. In the right panel, the characters of Mo atom $d$ orbitals are marked in different colors in the band. Without doping the Fermi level is at 0 eV.}
\end{figure}

Fixing the parameters, $\epsilon_{\text{d}} = 0.5$ eV, $U=5.0 $ eV, and $T=4.4$ K, our calculations yield the spin lifetime depending on $\mu$ as shown in the left panel of Fig.~\ref{fig3}. The right panel shows the TB band structure of MoS$_2$. The character of the bands are shown as "fat bands" in different color. For $0 ~\text{eV}<\mu<1.6$ eV, the spin lifetime is practically infinite, as there are no carriers in the MoS$_2$ bands for scattering. For $1.6$ eV $<\mu<1.9$ eV, only electrons from the $K$ valley can contribute to the spin scattering, which carry however mainly $d_{m=0}$ character. Thus, by symmetry we obtain low scattering rates and relatively long lifetimes. However, with chemical potentials in the range of $\mu<0$ eV or $\mu > 1.9$ eV electrons in MoS$_2$ with $d_{m=\pm2}$ orbital character contribute to efficient scatterings and short spin lifetime. Eventually, we arrive in the quantum regime with full Kondo screening of the impurity spin in these cases. Strikingly, as the Fermi level approaches the $\Sigma$ valley in the CB the spin lifetime drops extremely abrupt and by more than two orders of magnitude, further reaching the minimum ($\tau<10$ ps). Meanwhile, in the case of hole doping, $\mu<0$ eV, the spin lifetime decreases sharply to less than a nanosecond. Hence, information stored as "magnetic bit" in the adatom spin-orbital moment can be erased by tuning the electronic chemical potential $\mu$ in the MoS$_2$ sheet either to the VB or sufficiently high into the CB.

%\subsection*{SOC effect}

\begin{figure}
\includegraphics[width=\linewidth]{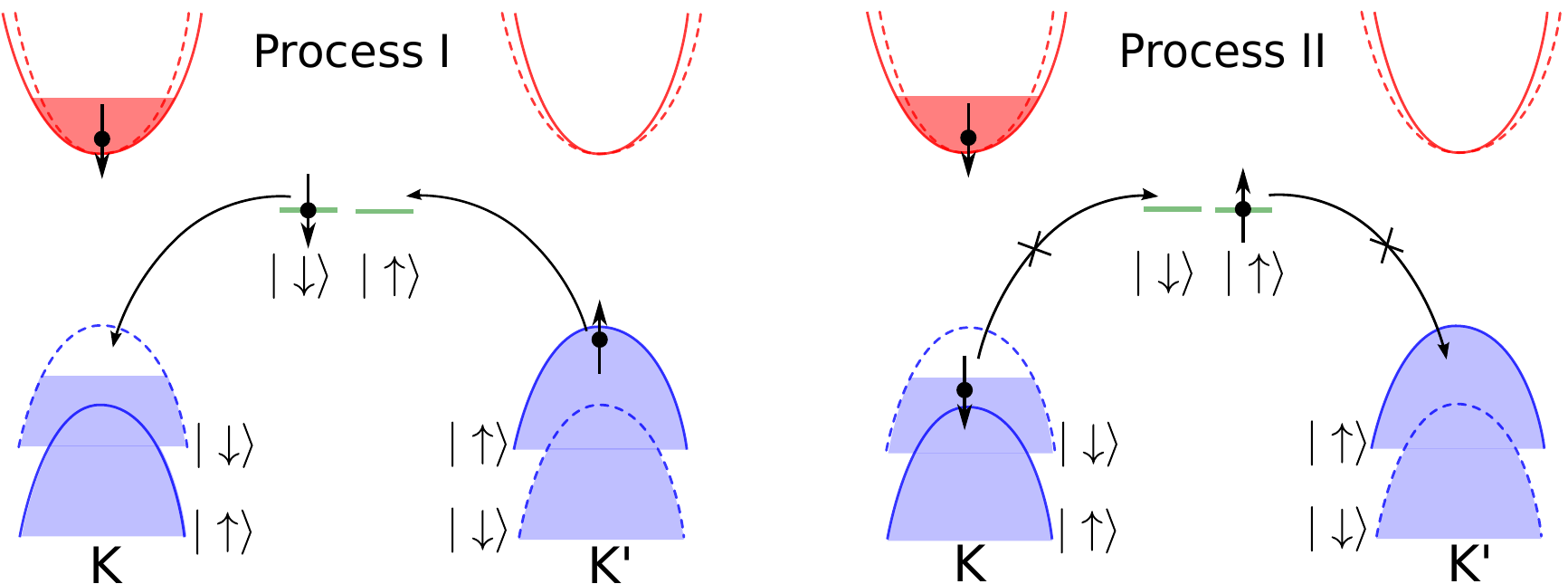}
\caption{\label{fig4}(Color online) Schematic illustration of the mechanism facilitating the optical manipulation of the spin-orbital moment. An optical pump with circularly polarized light excites the $\ket{\downarrow}$ states in the $K$ valley only. Due to the orbital selection rule, the excited electrons in the CB with mostly $d_{m=0}$ character can not efficiently scatter with the Co adatom. Process I: A spin $\ket{\downarrow}$ electron from one of the outmost Co adatom orbitals can scatter into one of the empty states of the MoS$_2$ VB at $K$ followed by a $\ket{\uparrow}$ electron from the VB at $K'$ hopping to the Co adatom. In this way the Co spin is flipped from $\ket{\downarrow}$ to $\ket{\uparrow}$. The reverse process II flipping the Co spin from $\ket{\uparrow}$ to $\ket{\downarrow}$ is not possible since no empty $\ket{\uparrow}$ state at $K'$ is available.}
\end{figure}

Due to the SOC in MoS$_2$, there are valley specific spin splittings of the CB in the $\Sigma$ valley and VB in the $K/K'$ valley~\cite{liu_three-band_2013} usually referred as spin-valley coupling~\cite{xiao_coupled_2012, suzuki_valley-dependent_2014,xu_spin_2014}. The variation of spin lifetimes obtained by including SOC also for the electrons of MoS$_2$ is shown as the red curve in the left panel of Fig.~\ref{fig3}. Because the variation of the spin lifetime is determined by the evolution of the orbital character of band electrons and SOC does not change this qualitatively, the curve with SOC follows the same trend of that without SOC. The major effect of SOC in the MoS$_2$ is the occurrence of "plateaus" in the $\tau$ vs. $\mu$ curve which can be attributed to the spin split bands in the $\Sigma$ and $K/K'$ valley, i.e., the spin-valley coupling.

Importantly, the spin-valley coupling provides a way to select the spin state of optically excited carriers in MoS$_2$ monolayers. To explain how this can be exploited for ultrafast optical orientation of the Co magnetic moment, we consider the resonant excitation of electrons from the highest spin-split VB to the lowest CB. In this case, the optically excited electrons in the CB mainly carry $d_{m=0}$ character so that due to the orbital selection rule their contributions to the scattering with the Co adatom are minor. Therefore, we only consider scattering with electrons in the VB with $d_{m=\pm2}$ character. The scattering can be still described by Eq.(\ref{eq7}), if the Fermi distribution functions $f(\epsilon)$ are replaced by the occupation numbers of the electron states in the VB after the optical excitation. To flip the Co spin, the spin of electrons from MoS$_2$ should be exchanged as well, i.e., the electrons in the VB of MoS$_2$ have to scatter between the $K$ and $K'$ valley. Specifically, the transition of an electron from MoS$_2$ from an occupied state in the $K'$ valley to an empty state in the $K$ valley flips the Co spin from $\ket{\downarrow}$ to $\ket{\uparrow}$ (process I in Fig.~\ref{fig4}). Its time reversed counterpart (process II in Fig.~\ref{fig4}) flips the Co spin from $\ket{\uparrow}$ to $\ket{\downarrow}$. Therefore, if the rate of process I is larger (smaller) than that of process II, the Co spin is written into a $\ket{\uparrow}$ ($\ket{\downarrow}$) state.

Using circularly polarized light, the excitation can be selectively done in the $K$ valley. Hence, we arrive at a situation, as shown in Fig.~\ref{fig4}, where $\ket{\downarrow}$ states in the $K$ valley are excited. At this moment, lacking empty states in the $K'$ valley, thus the process II is not allowed. In contrast, the process I can flip the Co spin from $\ket{\downarrow}$ to $\ket{\uparrow}$. As only one electron from the $K'$ valley will be transferred to the adatom, the number of empty states in the $K'$ valley remains at all times much smaller than the corresponding number in the $K$ valley. As long as this imbalance exists, the process II is blocked. Therefore, the adatom spin is written optically into a $\ket{\uparrow}$ state, while the recombination of electron-hole pairs in MoS$_2$~\cite{steinhoff_nonequilibrium_2016} will lead finally to an equilibrium without free carriers. We note that the same optical spin orientation mechanism also remains effective if the notoriously strong excitonic effects in materials like MoS$_2$ \cite{suzuki_valley-dependent_2014,xu_spin_2014} are accounted for.

In conclusion, we showed that orbital selection rules govern the spin-orbital dynamics of magnetic adatoms on a monolayer MX$_2$. Our study demonstrated that single Co, Ir and Rh adatoms on MoS$_2$ realize a $d^9$ valence state with a sizable magnetic anisotropy and highly orbital coupling to carriers in the substrate. This coupling lays the ground for electronic erasing and optical writing of information possibly stored as magnetic bits in the adatoms on monolayer TMDCs. Information processing and more generally optoelectronics based on valley degree of freedom have been appearing very promising but always feature the problem that valley degree of freedom entities like excitons are intrinsically volatile. The adatom spins could provide a means of storage and dynamic control of information encoded in the valley degree of freedom.

\section*{acknowledgements}
B.S. thanks the financial support from the fund of the BREMEN TRAC-COFUND Fellowship Programm of the University of Bremen. G.S. and M.S. acknowledge support via the Central Research Development Fund of the University of Bremen. The computations were performed with resources provided by the North-German Supercomputing Alliance (HLRN).


%\bibliography{Ising-TMDC}

\end{document}